\documentclass[showpacs,aps,prb,amsmath,amsfonts]{revtex4-1}
\usepackage{graphicx}
\usepackage{bm}
\newcommand{\mi}{\mathrm{i}}

\begin{document}

\title{Comment on ``Topological stability of the half-vortices
                    in spinor exciton-polariton condensates''}

\author{M. Toledo Solano}
 \affiliation{Centro de Investigaci\'on en Energ\'{\i}a, Universidad
Nacional Aut\'onoma de M\'exico, Temixco, Morelos, 62580, Mexico}
\author{Yuri G. Rubo}
 \email{ygr@cie.unam.mx}
 \affiliation{Centro de Investigaci\'on en Energ\'{\i}a,
Universidad Nacional Aut\'onoma de M\'exico, Temixco, Morelos, 62580, Mexico}

\date{February 11, 2010}

\begin{abstract}
We show that the conclusions of recent paper by Flayac \emph{et al}. [Phys.\
Rev.\ B \textbf{81}, 045318 (2010)] concerning the stability of half-quantum
vortices are misleading. We demonstrate the existence of static half-quantum
vortices in exciton-polariton condensates and calculate the warping of their
texture produced by TE-TM splitting of polariton band.
\end{abstract}

\pacs{71.36.+c,71.35.Lk,03.75.Mn}

\maketitle

Half-quantum vortices are topological excitations of multicomponent condensates
with combined spin-gauge symmetry \cite{VolovikBook}. They have been discussed
\cite{HVpap} to be the lowest energy topological excitations in
exciton-polariton condensates in semiconductor microcavities with unpinned
linear polarization of the condensate. The half-quantum vortices are expected
to define the Berezinskii-Kosterlitz-Thouless (BKT) transition in this
system,\cite{HVpap,Keeling08} and they have been recently observed
experimentally.\cite{LagoudakisScience}

In a recent paper\cite{Flayac} Flayac \emph{et al.} studied the effects of
TE-TM splitting of the exciton-polariton band on the state of vortices. The
authors of Ref.~\onlinecite{Flayac} have not found the static half-vortex
solutions of quasi-equilibrium two-component Gross-Pitaevskii equation (GPE),
and concluded that ``The half-vortices are no more stationary solutions of the
spinor Gross-Pitaevskii equations and should not affect the critical
temperature of the BKT phase transition". The goals of this Comment are to show
that this conclusion is incorrect, to indicate the mathematical error of
Ref.~\onlinecite{Flayac} that prevented the authors to establish the stationary
half-vortex solutions, and to present the correct way of solving this problem.

The order parameter of exciton-polariton condensate is the two-dimensional
complex vector $\bm{\psi}$. This vector can be written in terms of two
circular-polarization components, $\psi_{\pm}$, as
\begin{equation}
 \label{psicirc}
 \bm{\psi}
 =\frac{\hat{\mathbf{x}}+\mi\hat{\mathbf{y}}}{\sqrt{2}}\psi_{+}
 +\frac{\hat{\mathbf{x}}-\mi\hat{\mathbf{y}}}{\sqrt{2}}\psi_{-}.
\end{equation}
For the polariton condensate in quasi-equilibrium, the components of the order
parameter $\psi_{\pm}$ satisfy two coupled Gross-Pitaevskii equation [Eq.~(10)
of Ref.~\onlinecite{Flayac}]. In the case when TE-TM splitting is present, the
static vortex solutions can be found numerically, but one has to use the
correct asymptotic behavior of $\psi_{\pm}$. In Ref.~\onlinecite{Flayac} it was
assumed that the variables can be separated and, moreover, that the phases of
$\psi_{\pm}$ change linearly with the azimuthal angle [see Eq.~(12) of
Ref.~\onlinecite{Flayac}]. These assumptions are incorrect, and due to this
incorrect substitution (12) the static half-vortex solutions were not found in
Ref.~\onlinecite{Flayac}. In fact, we show below that \emph{the circular
phases} of the condensate order parameter \emph{are nonlinear functions of the
azimuthal angle}.

To find the asymptotic behavior of static solutions $\psi_{\pm}$ at large
distances $r$ from the half-vortex core (i.e., in the elastic region), we note
that in this region the polarization of the condensate becomes linear and the
amplitudes of the components of the order parameter becomes equal,
$|\psi_{\pm}|=\sqrt{n/2}$, where $n\equiv n(\infty)$ is the concentration of
the uniform condensate. Therefore, the asymptotics of $\psi_{\pm}$ can be
written as
\begin{equation}
 \label{Asymp}
 \psi_{\pm}(r\rightarrow\infty,\phi)=\sqrt{\frac{n}{2}}\,e^{\mi[\theta(\phi)\mp\eta(\phi)]}.
\end{equation}
Here we denoted the circular phases as $\theta\mp\eta$, using the common phase
angle $\theta$ and the polarization angle $\eta$, as defined in
Ref.~\onlinecite{HVpap}. These angles are yet unknown functions of the
azimuthal angle $\phi$.\cite{Notations}

To calculate the functions $\eta(\phi)$ and $\theta(\phi)$ we first obtain the
elastic energy of the condensate in the presence of TE-TM splitting.
Substitution of Eq.~(\ref{Asymp}) into the quasi-equilibrium Hamiltonian of
polariton condensate [Eqs.~(3-8) of Ref.~\onlinecite{Flayac}] gives
\begin{equation}
 \label{Hel}
 H_\mathrm{el}=\frac{\hbar^2n}{2m^*}\int dxdy
 \left\{
 (\nabla\eta)^2+(\nabla\theta)^2+
 2\gamma\left[
 \left(\frac{\partial e^{-\mi(\theta-\eta)}}{\partial z}\right)
 \left(\frac{\partial e^{ \mi(\theta+\eta)}}{\partial z}\right)
       +
 \left(\frac{\partial e^{ \mi(\theta-\eta)}}{\partial z^*}\right)
 \left(\frac{\partial e^{-\mi(\theta+\eta)}}{\partial z^*}\right)
 \right]
 \right\}.
\end{equation}
Here the effective mass $m^*$ and the TE-TM splitting parameter $\gamma$ are
defined by
\begin{equation}
 \label{gamma}
 \frac{1}{m^*}=\frac{1}{2}\left(\frac{1}{m_l}+\frac{1}{m_t}\right), \quad
 \gamma=\frac{m_t-m_l}{m_t+m_l},
\end{equation}
where the effective mass of transverse (or the transverse-electric, TE)
polaritons is $m_t$ and the effective mass of longitudinal (or
transverse-magnetic, TM) polaritons is $m_l$. In Eq.~(\ref{Hel}) we also use
the complex derivative
\begin{equation}
 \frac{\partial}{\partial z}=\frac{1}{2}
 \left(\frac{\partial}{\partial x}-\mi\frac{\partial}{\partial y}\right).
\end{equation}

The equations for the angles $\theta$ and $\eta$ are obtained by variation of
$H_\mathrm{el}$. In the limit $r\rightarrow\infty$ only azimuthal derivatives
are to be kept in these equations and we obtain
\begin{subequations}
 \label{Eqs_phi}
 \begin{equation}
 \left[1-\gamma\cos(2u)\right]\theta^{\prime\prime}+
 2\gamma\sin(2u)u^{\prime}\theta^{\prime}=0,
 \end{equation}
 \begin{equation}
 \left[1+\gamma\cos(2u)\right]u^{\prime\prime}
 +\gamma\sin(2u)\left( 1-u^{\prime2} -\theta^{\prime2} \right)=0,
 \end{equation}
\end{subequations}
where
\begin{equation}
 u(\phi)=\eta(\phi)-\phi.
\end{equation}

We note that the superfluid current $\mathbf{J}$ that corresponds to the
Hamiltonian density (\ref{Hel}) is given by more complicated expression than
the usual one. In particular, for the wave function (\ref{Asymp}) the radial
and azimuthal components of the current are
\begin{subequations}
 \label{Currents}
 \begin{equation}
 J_r=\frac{\hbar n}{m^*r}\gamma\sin(2u)\frac{d\theta}{d\phi},
 \end{equation}
 \begin{equation}
 J_\phi=\frac{\hbar n}{m^*r}\left[1-\gamma\cos(2u)\right]\frac{d\theta}{d\phi}.
 \end{equation}
\end{subequations}
The equation of continuity of the current for the static solutions,
$\mathrm{div}\mathbf{J}=0$, is reduced to the condition $dJ_\phi/d\phi=0$ and
gives Eq.~(\ref{Eqs_phi}a) again.

In what follows we denote the vortex solutions as $(k,m)$, with the
polarization and phase winding numbers $k$ and $m$, respectively. These numbers
are defined by $\eta(2\pi)-\eta(0)=2\pi k$ and $\theta(2\pi)-\theta(0)=2\pi m$,
and they can be integer or half-integer provided the sum $k+m$ is an
integer.\cite{HVpap} The topological charges used in Ref.~\onlinecite{Flayac}
are $l_{\pm}=m\mp k$. Eqs.~(\ref{Eqs_phi}a,b) have simple solutions for $k=1$
and an arbitrary integer $m$. Namely, $\eta=\phi$ and $\theta=m\phi$, so that
$u(\phi)\equiv0$. Only these solutions with $l_--l_+=2$ were claimed to exist
and were analyzed in Ref.~\onlinecite{Flayac}. The variables are indeed
separated in this case. Moreover, the polarization field for the $(1,0)$ vortex
(``hedgehog'') is purely longitudinal,\cite{HVString} and it is described by
the radial function of the usual vortex in one-component condensate with the
longitudinal mass $m_l$.

Apart from trivial solutions, Eqs.~(\ref{Eqs_phi}a,b) can be solved for any
other winding numbers $(k,m)$. Interestingly, there are two qualitatively
distinct type of solutions for elementary half-quantum vortices with
$k,m=\pm1/2$. In real microcavities the TE-TM splitting is small and we first
present the series of these solutions in the powers of $\gamma\ll1$. For the
$(1/2,\pm1/2)$ half-vortices the solutions are
\begin{subequations}
 \label{kplus12}
\begin{eqnarray}
 \theta(\phi)=\pm\left[
    \frac{1}{2}\phi+\frac{\gamma}{2}\sin(\phi)+\frac{\gamma^2}{4}\sin(2\phi)+\dots
                \right], \\
  \eta(\phi)=\frac{1}{2}\phi-\frac{\gamma}{2}\sin(\phi)+\frac{\gamma^2}{8}\sin(2\phi)+\dots\,.
\end{eqnarray}
\end{subequations}
For the $(-1/2,\pm1/2)$ half-vortices the solutions are
\begin{subequations}
 \label{kminus12}
\begin{eqnarray}
 \theta(\phi)=\pm\left[
    \frac{1}{2}\phi+\frac{\gamma}{6}\sin(3\phi)+\frac{\gamma^2}{36}\sin(6\phi)+\dots
                \right], \\
 \eta(\phi)=-\frac{1}{2}\phi+\frac{\gamma}{6}\sin(3\phi)-\frac{\gamma^2}{24}\sin(6\phi)+\dots\,.
\end{eqnarray}
\end{subequations}

The solutions to Eqs.~(\ref{Eqs_phi}a,b) can also be found numerically as shown
in Fig.~1. We have used a high value of the TE-TM splitting parameter,
$\gamma=0.3$, in order to illustrate the qualitative features of the behavior
of polarization and phase angles. Analyzing the superfluid current around the
vortex core we note that the streamlines are warped with respect to perfect
circles. Physically, the warping appears due to the change of the polariton
mass with polarization. Using Eqs.~(\ref{Currents}a,b) we find the streamlines
to be defined by equation
\begin{equation}
 \label{CurLine}
 \frac{d\mathrm{ln}r}{d\phi}=\frac{\gamma\sin[2u(\phi)]}{1-\gamma\cos[2u(\phi)]}.
\end{equation}
The warping of streamlines is shown in the inserts of Fig.~1.

The asymptotics found above divide the solutions of the Gross-Pitaevskii
equation into topologically distinct classes according to the values of winding
numbers $k$ and $m$. The half-vortex should be found by minimizing the full
Gross-Pitaevskii Hamiltonian within a particular topological class. This
solution exists since the energy is bound from below. This solution is static
since it is a minimum of the Hamiltonian. The half-vortices can be found either
by numerical solution of Gross-Pitaevskii equation with the required asymptotic
behavior, or by other means (e.g., by variational method).

Finally, we comment on the stability of half-vortices from the other point of
view. In the half-vortex core, when $r\rightarrow0$, one of the circular
components goes to zero and is singular: the order parameter behaves as
$re^{\pm\mi\phi}$ so that the gradient is not defined at $r=0$. For example,
for the $(1/2,1/2)$ half-vortex, $\psi_-\propto re^{\mi\phi}$ and
$\psi_+=\mathrm{const}$.\cite{Note0} If a half-vortex state is created by some
external means and evolves according to the time-dependent Gross-Pitaevskii
equation, this singularity in the solution will be present all the time. It is
because the Gross-Pitaevskii equation is regular and the vortex singularities
are rigid.\cite{NyeBerry} This is the reason why the estimation of life-time,
given by Eq.~(24) of Ref.~\onlinecite{Flayac} is not valid.

To conclude, we have shown that half-quantum vortices remain stable in the
presence of TE-TM splitting of polariton band, but their texture becomes
warped---the polarization and phase angles depend nonlinearly on the azimuthal
angle. We calculated the warping effect far away from the half-vortex core.

This work was supported in part by the grant No.\ IN112310 of DGAPA-UNAM.

\begin{figure}[t]
\includegraphics[width=4.0in]{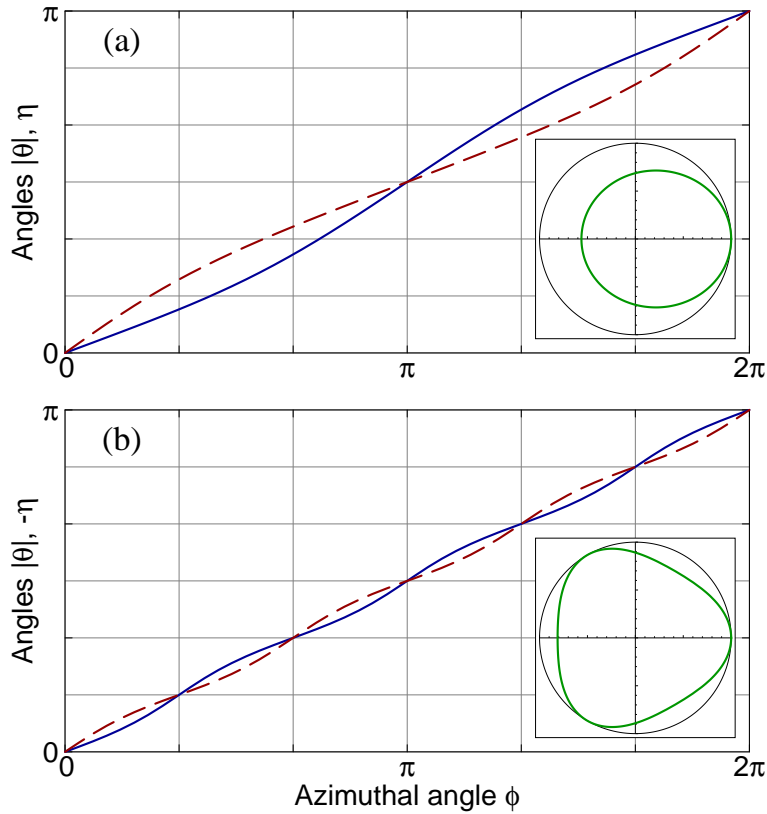}
\caption{\label{Fig:Warping} Showing the polarization angle $\eta(\phi)$ (solid blue line)
and the phase angle $\theta(\phi)$ (dashed red line) of the order parameter of
half-quantum vortices.
(a) The case of $(1/2,\pm1/2)$ half-vortices.
(b) The case of $(-1/2,\pm1/2)$ half-vortices.
Inserts show the warping of streamlines (thick green lines) of the current
around the half-vortex core with respect to the perfect circles (thin black lines).  }
\end{figure}

%%%%%%%%%%%%%%%%%%%%%%%%%%%%


\begin{thebibliography}{9}
\bibitem{Notations} Note that, contrary to Ref.~\onlinecite{Flayac},
    we write cylindrical coordinates as $(r,\phi)$. We use $\theta$
    to denote the common phase of the condensate order parameter.
\end{thebibliography}
\end{document}